\documentstyle[amstex,prd,aps,graphicx,epsf,tighten]{revtex}

\newcommand{\bra}{\left\langle}
\newcommand{\ket}{\right\rangle}

\newcommand{\bo}{{\mathcal{O}}}

\newcommand{\ltb}{LTB\ }
\newcommand{\vpc}{VPC}
\newcommand{\frw}{FLRW\ }

\begin{document}
\title{Averaging Spherically Symmetric Spacetimes in General Relativity}
\author{A.A. Coley\dag ~and N. Pelavas\dag}
\address{\dag\ Department of Mathematics and Statistics,\\
Dalhousie University, Halifax, Nova Scotia}

\maketitle

\begin{abstract}

We discuss the averaging problem in general relativity, using the
form of the macroscopic gravity equations in the case of spherical
symmetry in volume preserving coordinates. In particular, we
calculate the form of the correlation tensor under some reasonable
assumptions on the form for the inhomogeneous gravitational field
and matter distribution. On cosmological scales, the correlation
tensor in a Friedmann-Lema\^{\i}tre-Robertson-Walker (FLRW)
background is found to be of the form of a spatial curvature. On
astrophysical scales the correlation tensor can be interpreted as
the sum of a spatial curvature and an anisotropic fluid. We
briefly discuss the physical implications of these results.

\end{abstract}

\pacs{98.80.Jk,04.50.+h} \noindent [PACES: 98.80.Jk,04.50.+h] \vskip1pc

The gravitational field equations on large scales are obtained by
averaging the Einstein equations of general relativity (GR). The
Universe is not isotropic or spatially homogeneous on local
scales. An averaging of inhomogeneous spacetimes on large scales
can lead to important effects. For example, on cosmological scales
the dynamical behavior can differ from that in the spatially
homogeneous and isotropic Friedmann-Lema\^{\i}tre-Robertson-Walker
(FLRW) model \cite{Ellis:1984}; in particular, the expansion rate
may be significantly affected. Consequently, a solution of the
averaging problem is of considerable importance for the correct
interpretation of cosmological data. It is also of importance for
physical phenomena on  astrophysical (galactic) scales.

There are a number of theoretical approaches to the averaging
problem \cite{Bild-Futa:1991,Zala,buch}. In the approach of
Buchert \cite{buch}  a 3+1 cosmological space-time splitting is
employed and only scalar quantities are averaged. The perturbative
approach \cite{Bild-Futa:1991} involves averaging the perturbed
Einstein equations; however, a perturbation analysis cannot
provide any information about an averaged geometry. On the other
hand, the macroscopic gravity (MG)  approach to the averaging
problem in GR \cite{Zala} gives a prescription for the correlation
functions which emerge in an averaging of field equations. The MG
approach is a fully covariant, gauge independent and exact method.
We shall adopt the  MG averaging approach. Averaging of the
structure equations for the geometry of GR then leads to the
structure equations for the averaged (macroscopic) geometry and
the definitions and the properties of the correlation tensor. The
averaged Einstein equations can always be written in the form of
the Einstein equations for the macroscopic metric tensor when the
correlation terms are moved to the right-hand side of the averaged
Einstein equations \cite{Zala}.

Spherical symmetry is of particular physical interest, and it is
especially important to study the averaging problem within the
class of spherically symmetric cosmological models. In
\cite{CPpaper} the microscopic field equations were taken and the
averaging procedure was effected to determine the precise form of
the correlation tensor in this case. In volume preserving
coordinates (VPC), the spherically symmetric line element is given
by

\begin{equation}
ds^2=-Bdt^2+Adr^2+\frac{du^2}{\sqrt{AB}(1-u^2)}+\frac{1-u^2}{\sqrt{AB}}d\phi^2,
\label{vpcss}
\end{equation}

\noindent where the functions $A$ and $B$ depend on $t$ and $r$.
The FLRW metric in VPC is given by (\ref{vpcss}), with $A  =
{R^2}/{F^4}, B  =  {1}/{R^6}$, where $R= R(t)$ and $F=F(r)$,
subject to $ \frac{dF}{dr} = {\sqrt{1-kF^2}}/{F^2} $ and $k=-1,0$
or $1$. We can calculate the form of the Einstein tensor $G^{a}_{\
b}$, take averages, and obtain the appropriate form for the MG
field equations and hence the correlation tensor $C^{a}_{\ b}$
(for example, we have that $C^{r}_{\ t}=G^{r}_{\ t}\bra g \ket -
\bra G^{r}_{\ t}\ket$). In VPC, the average is then simply given
by

\begin{equation}
\label{bra}
\bra f(r,t) \ket \equiv
\frac{1}{TL}\int^{\frac{T}{2}}_{t^{\prime}=-\frac{T}{2}}
dt^{\prime} \int^{\frac{L}{2}}_{r^{\prime}=-\frac{L}{2}}
dr^{\prime}f(r+r^{\prime}, t+t^{\prime}),
\end{equation}

\noindent which, for smooth functions with a slowly varying
dependence on cosmological time, essentially reduces to a spatial
average in terms of the averaging scale  $L$ (with $L \equiv h_0/H
< 1$).

The form of the correlation tensor depends on the assumed form for
the inhomogeneous gravitational field and matter distribution. We
assume that
\begin{equation}
\label{Matter}
A(r,t)=\bra A(r,t) \ket \left[ 1+
\sum_{n=1}^{\infty}a_{n}(t)L^{n}\sin\left(\frac{2n\pi}{L}r\right)
+
\sum_{n=1}^{\infty}{\bar{a}}_{n}(t)L^{n}\cos\left(\frac{2n\pi}{L}r\right)
\right],
\end{equation}
\begin{equation}
\label{Inhom} B(r,t)=\bra B(r,t) \ket \left[ 1+
\sum_{n=1}^{\infty}b_{n}(t)L^{n}\sin\left(\frac{2n\pi}{L}r\right)
+ \sum_{n=1}^{\infty}{\bar
{b}}_{n}(t)L^{n}\cos\left(\frac{2n\pi}{L}r\right) \right],
\end{equation}
where the  inhomogeneous functions $A$ and $B$ satisfy a set of
appropriate and self-consistent conditions (for example, $ \bra
\frac{\partial}{\partial t}\bra A(r,t) \ket \ket=\bra
\frac{\partial A(r,t)}{\partial t} \ket$). Expanding in powers of
$L <1$, we obtain the correlation tensor up to $\bo(L^2)$
\cite{CPpaper}:

\begin{equation}
\label{star} C^a \; _b = diag\left[C+ \frac{2 \ell}{\langle A
\rangle}, C, \frac{\ell}{\langle A \rangle},  \frac{\ell}{\langle
A \rangle}\right],
\end{equation}
where $C \equiv C^r \;_r$ and
$$ \ell(t) \equiv \frac{\pi^2}{8}\left[(a_1 -3b_1)(a_1 +b_1)+(\bar{a}_1
-3\bar{b}_1)(\bar{a}_1 +\bar{b}_1) \right]. $$ The function $C$
can then be calculated from the contracted Bianchi identities.  We
note that if $C^a \; _b$ is isotropic (i.e., of the form of a
perfect fluid) then $C = \frac{\ell}{\langle A \rangle}$ and $C^a
\; _b$ is of the form of a spatial curvature term. Hereafter, for
convenience we shall drop the angled brackets on averaged
quantities.

Let us first discuss averaging on cosmological scales. In the case
that $\frac{\partial B}{\partial r} =0$, as in the case of a FLRW
background, the contracted Bianchi identities immediately yield $C
\equiv \ell/ A$ and $A _{, r} =0$, and $\ell / A = \ell_0 R^{-2}$
(where $\ell_{0}$ is a constant). Therefore, in this case we
obtain
\begin{equation}
\label{stars}
 C^a \; _b = \ell_0 R^{-2}diag\left[3,1,1,1\right] \; ,
\end{equation}
and $C^a \; _b$ is necessarily of the form of a spatial curvature
term.

The cosmological result that in the spherically symmetric case the
averaged Einstein equations in an FLRW background have the form of
the Einstein equations of GR for a spatially homogeneous,
isotropic macroscopic space-time geometry with an additional
spatial curvature term, confirms the results in previous work in
which we were able to explicitly solve the MG equations to find a
correction term (correlation tensor) in the form of a spatial
curvature \cite{CPZ}. This result is also (i) consistent with the
work of Buchert \cite{buch}, in which a spatial curvature term
appears when averaging in a FLRW background, (ii) consistent with
the results of averaging an exact Lema\^{\i}tre-Tolman-Bondi (LTB)
spherically symmetric dust model \cite{LTBAV}, in which solutions
of the \ltb metric in (non-diagonal) \vpc\  are given explicitly
as perturbations about the spatially flat \frw model and found to
give rise to solutions which can be interpreted as having both
spatial curvature and a constant correction term, and (iii)
consistent with results in which the effects of linear
inhomogeneous perturbations perturbations on an exact spatially
homogeneous and isotropic FLRW background
\cite{Bild-Futa:1991,AV1} are found to give rise to correlation
terms of the form of a spatial curvature term.

Inhomogeneities can affect the dynamics and may significantly
affect the expansion rate of the spatially averaged ``background''
FLRW universe (the effect depending on the scale of the initial
inhomogeneity) \cite{Bild-Futa:1991}. Therefore, a more
conservative approach to explain the acceleration of the Universe
\cite{SN} without introduction of exotic fields is to utilize a
back-reaction effect due to inhomogeneities of the Universe.
Indeed, it has been suggested that back-reactions from
inhomogeneities could explain the apparently observed accelerated
expansion of the universe today. This has been investigated by
studying the effective Friedmann equation describing an
inhomogeneous Universe after averaging, using both perturbative
and qualitative analyses \cite{AV1,AV2}. It is clear that the
perturbative effect proposed always gives rise to a
renormalisation of the spatial curvature. It has also been argued
that the effect does not simply reduce to spatial curvature and an
acceleration can also result (although it is unlikely to be
compatible with other observational data).

The MG method adopted here is an exact approach in which
inhomogeneities affect the dynamics on large scales through the
correlation term. Averaging can have a very significant dynamical
effect on the evolution of the Universe; the correction terms
change the interpretation of observations so that they need to be
accounted for carefully to determine if the models may be
consistent with an accelerating Universe. Averaging may or may not
explain the observed acceleration. However, it is clear that it
cannot be neglected, and a proper analysis will not be possible
without a comprehensive understanding of the affects of averaging.

Let us next consider the effects of averaging on astrophysical
scales (e.g.,  galactic scales). We assume that a galaxy can be
approximated as spherically symmetric. In a non-FLRW background
(with $\frac{\partial B}{\partial r} \neq 0$), the contracted
Bianchi identities  can then be integrated to obtain
\cite{CPpaper}
 \begin{equation}
 C = -\frac{\ell}{A} + f(t)
 \frac{(AB)^{1/2}}{A^{2\ell}}~~;~~~~~
 \dot{\ell} = - \left[ \frac{2f}{A^{2 \ell}}  \right]_{, t}
 A^{3/2} B^{1/2}.
 \label{dddag}
  \end{equation}
We note that $C^a \; _b$ is necessarily anisotropic (and cannot be
formally equivalent to a perfect fluid). For the solution with
${\ell}={\ell}_0=const.$ and   $2f = g(r)A^{2 \ell}$, we can
always write
\begin{equation}
C^a \; _b = \ell_0 A^{-1} diag\left[3,1,1,1 \right] -\Pi
diag\left[1,1,0,0\right] \label{eqn.s}
\end{equation}
where $\Pi \equiv -\{g(r) {AB}^{1/2} - 2 \ell_0  A^{-1} \}$. The
correlation tensor $C^a \; _b$ then automatically satisfies the
contracted Bianchi identities. It can be interpreted as the sum of
a perfect fluid and an anisotropic fluid (when $B_r \ne 0$). If
both terms separately satisfy the contracted Bianchi identities,
then the first term can be interpreted as a spatial curvature term
and the second term can be interpreted as an anisotropic fluid
with $p_\perp = 0$ and $p_{||} = - \rho_{\mbox{eff}}$. For an
anisotropic fluid in spherically symmetric models the
energy-momentum tensor is of the form $diag\left[ -\mu, p_{||},
p_\perp, p_\perp\right]$, where $p_{||} = p + \frac{2}{3}\pi$ and
$p_\perp = p -\frac{1}{3}\pi$, and $\pi$ is the anisotropic
pressure. From above, we see that if the (total) correlation
tensor $C^a \; _b$ is interpreted as an anisotropic fluid (which
is comoving in VPC), it follows that $\Pi = - \pi$ and $p = -
\frac{1}{3}\mu$. Anisotropic fluids in spherically symmetric
models have been studied in \cite{aniso}.

Although the correlation tensor $C^a \; _b$ satisfies the
contracted Bianchi identities, when interpreted as the sum of a
spatial curvature perfect fluid and an anisotropic fluid through
(\ref{eqn.s}), the two separate fluid do not in general satisfy
separate conservation equations. However, the contracted Bianchi
identities can be rewritten in the form of a conservation law for
the anisotropic pressure $\pi$,

\begin{equation}
\label{cons} \pi_t -\frac{1}{2} \pi \left( \frac{A_t}{A} -
\frac{B_t}{B} \right) + \frac{\ell_0}{A} \left(2\frac{A_t}{A} +
\frac{B_t}{B}  \right) =0,
\end{equation}
in VPC where the metric is given by eqn. (\ref{vpcss}).

Let us comment on the astrophysical applications of an anisotropic
fluid. It is known that dark matter is a major constituent of the
halos of galaxies \cite{DM}. By an analysis of observed rotation
curves, under reasonable assumptions (e.g., that galaxies can be
modelled as spherically symmetric) it has been found that the dark
matter is of the form of an anisotropic fluid \cite{Lake}. This
has been taken up in \cite{Boon}, in which the consequences of
anisotropic dark matter stresses are discussed in the weak field
gravitational lensing (where it was noted that any attempt to
model dark matter in galactic halos with classical fields will
lead to anisotropic stresses comparable in magnitude with the
energy density).

It is of interest to further study the effects of averaging in the
astrophysical context. The results of this work could be used to
model the effects phenomenologically by including an anisotropy
term (comoving in VPC), which in general has $p = -
\frac{1}{3}\mu$, and in the case $\ell_0=0$ is of the specific
form $p_\perp = 0$ and $p_{||} = - \rho_{\mbox{eff}}$ (where the
correlation tensor is given by $-\Pi diag\left[1,1,0,0\right]$).
The anisotropic fluid satisfies the Bianchi identities, but since
it arises from an averaging procedure it need not satisfy any
energy conditions. It may be beneficial to work in VPC, in which
the metric is diagonal and the correlation tensor is `comoving'
(although the matter is not generally comoving). Indeed, in VPC
the correlation tensor is given explicitly in terms of the
averaged metric functions (e.g., $\mu = -g(r) {AB}^{1/2} -  \ell_0
A^{-1}$, $p = - \frac{1}{3}\mu$, $\pi = g(r) {AB}^{1/2} - 2 \ell_0
A^{-1}$). A disadvantage is that astrophysicists are not familiar
with working in these coordinates. Alternatively, we could
transform back to more conventional coordinates and determine the
form of the correlation tensor; however, these coordinates may not
be the most natural (e.g., the metric will not be diagonal) and
the form of the correlation tensor (which is no longer comoving)
may be quite complicated.

{\em Acknowledgements}. This work was supported, in part, by
NSERC.

\end{document}